\documentclass[aps,prl,reprint,groupedaddress]{revtex4-2}

\usepackage{graphicx}
\usepackage{color}
\usepackage{amsmath}
\usepackage{hyperref}

\newcommand{\HgT}{HgBa$_2$Ca$_2$Cu$_3$O$_{8+ \delta}$~}
\newcommand{\BiT}{Bi$_2$Sr$_2$Ca$_2$Cu$_3$O$_{10+ \delta}$~}
\newcommand{\YBCO}{$\mathrm{YBa_2Cu_3O_{6+x}}$}

\newcommand{\bA}{\mathbf{A}}
\newcommand{\QQ}{\mathbf{Q}}

\newcommand{\bq}{\mathbf{q}}

\newcommand{\br}{\mathbf{r}}
\newcommand{\bk}{\mathbf{k}}

\begin{document}

\title{Charge order near the antiferromagnetic quantum critical point in the trilayer high $T_c$ cuprate \HgT}

\author{V. Oliviero$^{1,\dagger}$, I. Gilmutdinov$^{1,\dagger}$, D. Vignolles$^{1}$, S. Benhabib$^{1,\parallel}$, N.~Bruyant$^{1}$, A. Forget$^{2}$, D. Colson$^{2}$, W.A. Atkinson $^{3,*}$ and C. Proust$^{1,\ddagger}$}

\affiliation{
$^1$LNCMI-EMFL, CNRS UPR3228, Univ. Grenoble Alpes, Univ. Toulouse, INSA-T, Grenoble and Toulouse, France.\\ 
$^2$Service de Physique de l’Etat Condensé, CEA Saclay (CNRS-URA 2464), Gif sur Yvette 91191, France.\\
$^3$Department of Physics and Astronomy, Trent University, Peterborough, K9L 0G2, Ontario, Canada.\\
}

\date{\today}

\begin{abstract}
We study the transport properties of underdoped trilayer cuprate \HgT with doping level $p$ = 0.1 - 0.12 in magnetic field up to 88~T. We report for the first time in a cuprate superconductor a dramatic change of the quantum oscillation spectrum versus temperature, which is accompanied by a sign change of the Hall effect below $T \approx$ 10~K. Based on numerical simulations, we infer a Fermi surface reconstruction in the inner plane from an antiferromagnetic state (hole pockets) to a biaxial charge density wave state (electron pockets). We show that both orders compete and share the same hotspots of the Fermi surface and we discuss  our result in the context of spin-fermion models.
\end{abstract}

\maketitle

One of the surprising features of cuprate superconductors is the ubiquity of the interplay between antiferromagnetic (AFM) order, charge density waves (CDWs), and superconductivity across cuprate families \cite{Keimer15}.  In general, these orders compete and each occupies its own piece of the phase diagram; however, significant coexistence regimes are  observed and these can lead to novel cooperative behavior.  As a prominent example, incommensurate CDW and spin orders coexist as magnetic stripes in the La-based cuprates \cite{Frano20,Uchida21}. 
In other cuprates, CDWs are nonmagnetic and there have been ongoing efforts to understand their connection to the strong electron correlations, low dimensionality, and AFM fluctuations that are hallmarks of the cuprates \cite{Keimer15}.
The Peierls paradigm that requires extended sections of Fermi surface to be nested by some wavevector ${\bf q}$ is certainly not at the origin of the CDW formation in cuprates. There are other materials exhibiting CDWs without obvious Fermi surface nesting---for example, layered dichalcogenides such as NbSe$_2$ \cite{zhu_misconceptions_2017}, nonmagnetic pnictides in the BaNi$_2$As$_2$ family \cite{Lee:2019pnictide}--- for which there is compelling evidence that the CDWs are phonon-driven \cite{johannes_fermi_2008,Weber:2013NbSe2,Souliou:2022soft,Song:2023phonon}. The corresponding evidence in the cuprates is weak \cite{Souliou:2021YBCO}.

As an alternative to phonons, spin fluctuations provide a natural mechanism for CDW formation in cuprates. The mechanism for high temperature superconductivity in the cuprate is widely accepted to be of magnetic origin \cite{Scalapino:2012RMP}, and several analytical calculations predict that spin-fluctuation-mediated CDWs are degenerate with superconductivity at the AFM quantum critical point (where the N\'eel temperature vanishes) \cite{Metlitski:2010quantum,Metlitski:2010instabilities,Efetov:pseudogap2013,Wang:2014cdw}.  This result is important because it provides a route to charge order at high temperatures without Fermi surface nesting. Instead, CDW wavevectors connect Fermi surface ``hotspots'' that are determined by the ${\bf q}$-dependence of the spin susceptibility.  Early versions of the theory incorrectly predicted that the CDW ${\bf q}$-vectors lie along the Brillouin zone diagonals, with ${\bf q} = (q,\pm q)$; however, quantitatively correct axial CDW wavevectors, ${\bf q} = (q,0)$, $(0,q)$, are obtained when the dominant diagonal-${\bf q}$ instability is suppressed.  This occurs close to the AFM quantum critical point \cite{Pepin:pseudogap_2014,Allais:2014density,Tsuchiizu:2016}, within  AFM  \cite{Wang:2018fragility,Atkinson:charge2015} or loop-current  \cite{Atkinson_emergence_2016} phases, or when strong correlations are accounted for \cite{Chowdhury:2014,banerjee2020emergent,Banerjee:2022,mascot2022electronic}. It is important to point out, however, that realistic band structures produce weak peaks in the bare charge susceptibility with  similar ${\bf q}$-vectors \cite{sachdev_bond_2013} despite the lack of proper Fermi surface nesting. 
Thus, while there is  experimental support for the role of Fermi surface hotspots \cite{atkinson_structure_2018}, it remains an open question as to whether they are generated by the Fermi surface or the interaction.
\begin{figure*} [t]
\includegraphics[width=1\textwidth]{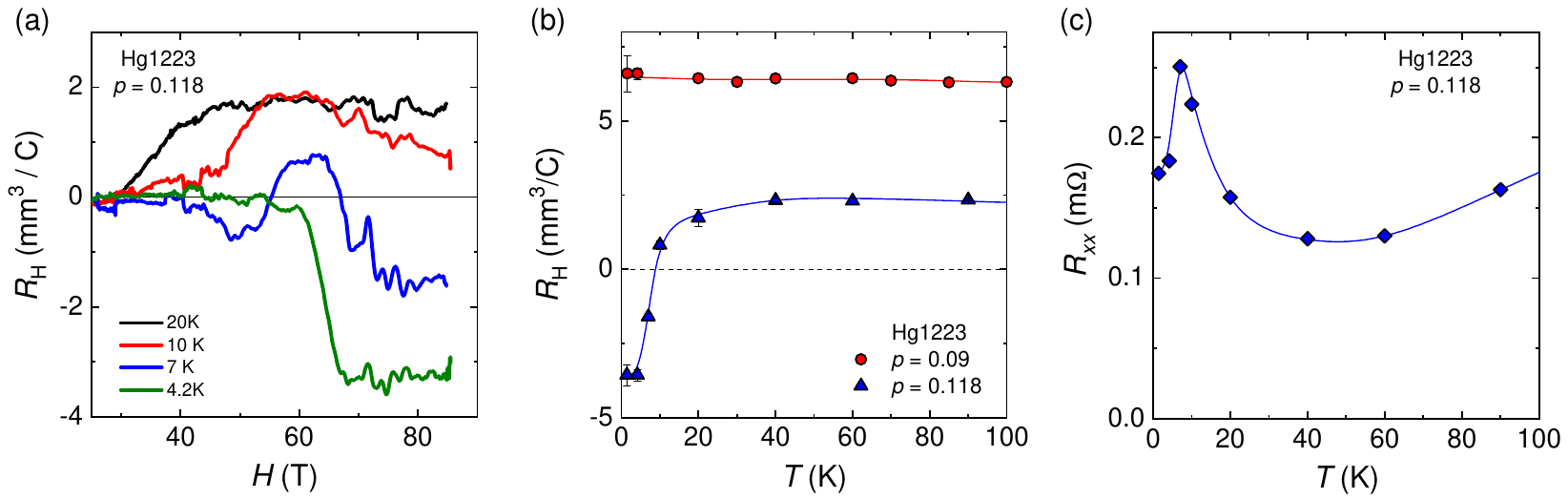}
\caption{\textbf{} a) Field dependence of the Hall coefficient $R_H$ in Hg1223 ($p$ = 0.118) at various fixed temperatures, as indicated.
b) Temperature dependence of the normal-state Hall coefficient $R_H$, measured at high fields, in Hg1223 $p$ = 0.09 (red circles from \cite{Oliviero22} and $p$ = 0.118 (blue triangles, this work). At $p$ = 0.118, $R_H$ changes sign abruptly below $T$ = 10~K  while it remains positive down to the lowest temperature for $p$ = 0.09.
c) Temperature dependence of the normal-state resistance in Hg1223 ($p$ = 0.118). The sign change of the Hall coefficient is accompanied by a maximum in the resistance, that strongly suggest the occurrence of a phase transition.  
}
\label{Fig1}
\end{figure*}

CDWs have clear signatures in the transport properties of \YBCO~(YBCO) and single-layer HgBa$_2$CuO$_{4+ \delta}$ (Hg1201). 
At low temperature, the observation of small-frequency quantum oscillations \cite{Doiron07,Barisic13} combined with negative Hall \cite{LeBoeuf07,LeBoeuf11,Doiron13} and Seebeck \cite{Chang2010,Laliberte11,Doiron13} coefficients indicate the presence of a closed electron pocket in the reconstructed Fermi surface. 
The exact mechanism of Fermi surface reconstruction by the CDW in underdoped cuprates is still debated \cite{Proust19} but a plausible scenario is a biaxial charge order that creates a small electron-like pocket located at the nodes. Quantum oscillations with small frequencies have also been observed in the underdoped trilayer cuprate \HgT (Hg1223) at hole doping $p$ = 0.08-0.09 but without a sign change in the Hall coefficient down to the lowest temperatures \cite{Oliviero22}. A plausible interpretation is the coexistence of an AFM order in the inner plane and CDW in the outer planes. Multilayer cuprates, with three or more CuO$_2$ layers per unit cell, offer a new twist on the story because a single material can host distinct phases in different layers simultaneously \cite{Mukuda12}. This raises intriguing questions about the coexistence and competition between these phases, and the possible emergence of novel cooperative phases. In particular, AFM order is more robust in multilayer cuprates and persists up to higher doping as the number of CuO$_2$ planes in the structure increases \cite{Mukuda12}. The symmetry-inequivalent CuO$_2$ planes gives rise to a disorder-protected inner plane and the fact that the outer layers are closer to charge reservoir gives rise to a charge imbalance \cite{Mukuda12,Ideta10,Kunisada20,Luo22}.

In this Letter, we explore CDW formation near the AFM quantum critical point in the trilayer cuprate Hg1223. 
We performed quantum oscillation and Hall effect measurements in pulsed magnetic fields up to 88~T and observed clear signatures of a Fermi surface reconstruction below 10~K. Through numerical modeling, we attribute the temperature evolution of the quantum oscillation spectrum to competition between weak antiferromagnetism and charge order, leading to a sequence of crossovers as the temperature is reduced.  Consistent with spin-fluctuation models, we find that the phase competition occurs because the CDW and AFM share hotspots.

Quantum oscillation measurements using the tunnel diode oscillator (TDO) technique \cite{Coffey00, Drigo10} were performed at the pulsed-field facility in Toulouse (LNCMI-T) up to 88~T in two samples of Hg1223 at doping levels $p=0.102$ ($T_c = 96$~K) and $p=0.112$ ($T_c = 108$~K); (see Supplemental Information for more details). The doping level has been estimated from $T_c$ and represents an average of the inner and  outer layers. The Hall effect was measured in two distinct samples of Hg1223 at similar doping levels, $p=0.101$ ($T_c = 95$~K) and $p=0.118$ ($T_c = 114$~K);. The magnetic field $H$ was applied along the $c$-axis of the tetragonal structure, perpendicular to the CuO$_2$ planes (both field polarities for the Hall effect).

Figure~\ref{Fig1}(a) shows the Hall coefficient plotted as a function of magnetic field $H$ for $p = 0.118$ at different temperatures from $T = 20$~K down to $T = 4.2$~K. While $R_H$ is positive above $T = 20$~K, it becomes negative at $T = 4.2$~K and flattens at fields above $H \approx 70$~T, which indicates that the normal state is reached. The temperature-dependence of the normal-state Hall coefficient measured at the highest fields between $T$ = 1.5~K and 100~K is shown in Fig.~\ref{Fig1}(b) at two doping levels, $p = 0.118$ (this work) and $p = 0.09$ \cite{Oliviero22}. For $p = 0.118$, there is a sudden sign change of the Hall coefficient below $T \approx$~ 10~K that is not observed at lower doping. The sign change is accompanied by a maximum of the resistance at $T \approx$~10~K [Fig.~\ref{Fig1}(c)].  In other cuprates, notably underdoped YBCO \cite{Wu11, Ghiringhelli12, Chang12} and Hg1201 \cite{Tabis14,Chan20}, this behaviour is attributed to the emergence of charge order. Here, both the abruptness of the transition and its low temperature, are surprising.

\begin{figure*} [t]
\includegraphics[width=1\textwidth]{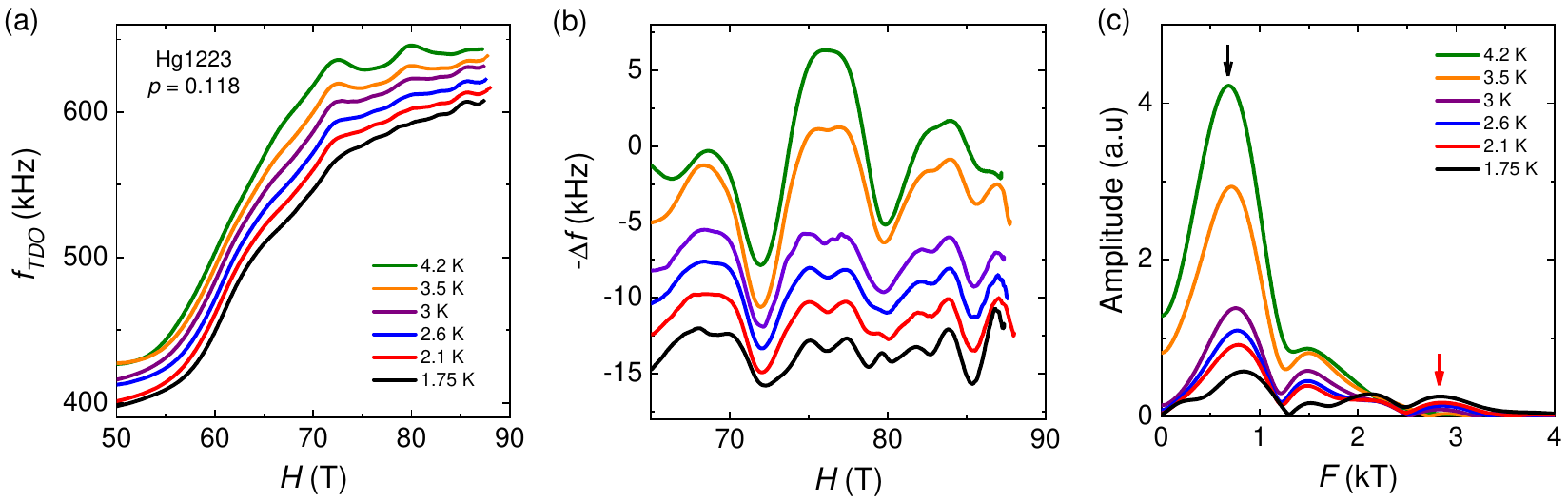}
\caption{\textbf{} a)  Field dependence of the TDO frequency after the heterodyne circuit in  Hg1223 ($p$ = 0.11) at various fixed temperatures, as indicated. 
b) Oscillatory part of the TDO signal after removing a smooth background (spline) from the data shown in panel a). -$\Delta f$ is plotted and the curves have been offset for clarity.
c) Discrete Fourier analysis of the oscillatory part of the TDO signal shown in panel b). The black (red) arrow marks the low (high) frequency observed at $T$ = 4.2~K ($T$ = 1.8~K). 
}
\label{Fig2}
\end{figure*}
Because of the low temperatures, we are in the rare situation of being able to track the Fermi surface morphology through the CDW transition using quantum oscillation measurements. Figure~\ref{Fig2}(a) shows 
the TDO frequency (see SI) as a function of magnetic field at different temperatures for the $p=0.118$ sample. A smooth background subtraction leads to the oscillatory part of the signal depicted in Fig.~\ref{Fig2}(b).
At $T=4.2$~K, there is a strong low-frequency oscillation whose amplitude decreases with decreasing temperature.   By $T=1.75$~K, the low-frequency oscillations are weak, and small-amplitude oscillations at higher frequencies have emerged.
This is inconsistent with the Lifshitz-Kosevich theory for a temperature-independent Fermi surface \cite{Shoenberg}, but rather signals a Fermi surface reconstruction. 
The temperature evolution of the oscillation spectrum is clearly seen in the Fourier analysis of the oscillatory part of the data  (Fig.~\ref{Fig2}(c)). At 4.2~K, the spectrum is dominated by a low-frequency peak ($F = 680$~T, black arrow) and its first harmonic. The low-frequency peak weakens and shifts to higher frequency ($F=830$~T) as the temperature decreases. Two peaks at high frequency, $F = 2100$~T and $F = 2800$~T (red arrow), emerge at the lowest temperatures.  While the high frequencies could result from a large effective mass (a Lifshitz-Kosevich analysis for $F = 2800$~T gives $m^* \approx 6~m_e$), the Fermi surface reconstruction scenario is required by the $T$-dependence of the low-frequency oscillation. Note that both signatures of a Fermi surface reconstruction in Hg1223---the sign change of the Hall coefficient and the evolution of the quantum oscillation spectrum versus temperature---have been reproduced in other samples at doping level $p = 0.10$ (see SI).\\

\begin{figure*}
	\includegraphics[width=\textwidth]{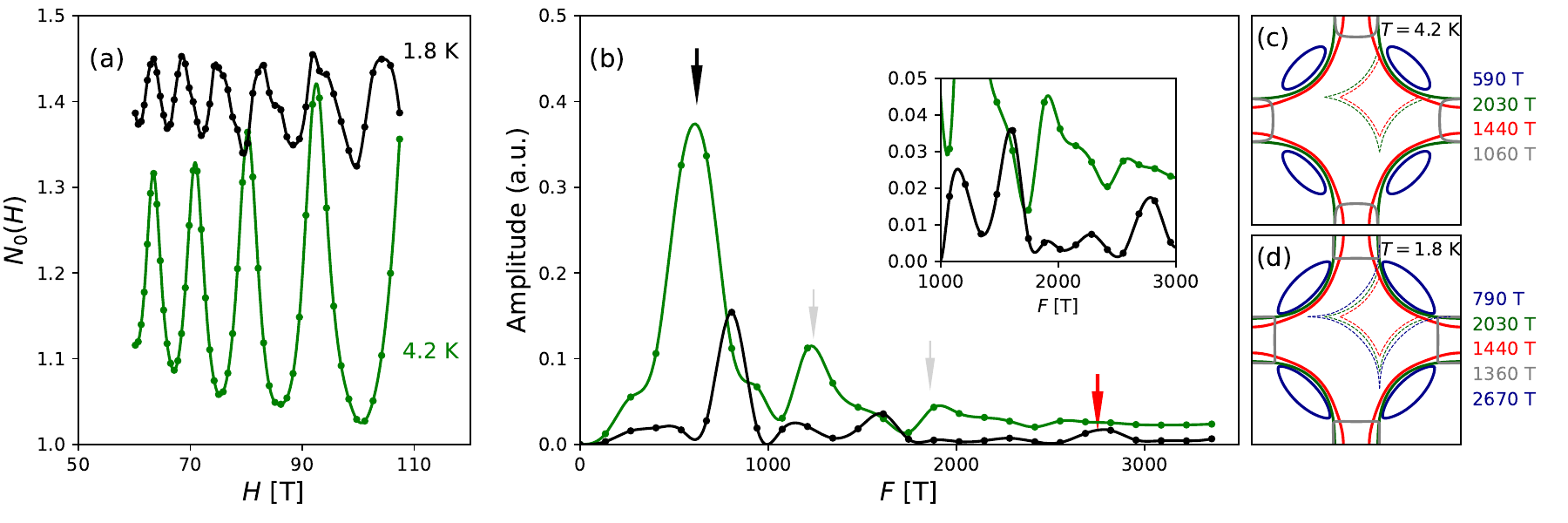}
    \caption{(a) Calculated density of states $N_0(H)$ at the Fermi energy as a function of magnetic field $H$ for $T=4.2$~K and $T=1.8$~K and (b) its Fourier transform, using the parameters in Table~\ref{Table1}.  
    The inset shows an expanded view of the high frequency region. Arrows indicate the low- (black) and high-frequency (red) peaks and their harmonics (grey). (c) and (d) show simplified Fermi surfaces for the 4.2~K and 1.8~K, respectively.  Quantum oscillation frequencies calculated from the Fermi surface areas are listed next to each panel. 
    }
    \label{Fig_theo}
\end{figure*}

Our results demonstrate that a crossover takes place at low temperature in the doping range $p$ = 0.10 - 0.12, when superconductivity is quenched by a magnetic field. To gain insight, we numerically simulated quantum oscillations of the density of states for a single CuO$_2$ trilayer (see SI).  We found that  CDW order alone can not explain the measured oscillations; however, we obtain quantitatively correct oscillation frequencies if we assume that there is weak AFM order in the inner layer. The required AFM order parameter $M$ is quite small ($M \lesssim 50$~meV) suggesting that the system is close the AFM quantum critical point. We introduce distinct CDW order parameters $\Delta_i$ and $\Delta_o$ on inner and outer layers, respectively. To obtain closed semiclassical orbits, we assume that the CDWs are  biaxial, with wavevectors ${\bf q}_a = (q, 0)$ and $\bq_b= (0, q)$ where $q=2\pi/4.5a_0$, and with a bond-centered $d$-wave form factor.  The true form factor is likely more complex \cite{McMahon20}, but would not change our results qualitatively.  Conversely, our choice of $q$ is essential to obtain the correct oscillation frequencies at low $T$.  We include anisotropic scattering rates $\Gamma_{o,\bf k}$ and $\Gamma_{i,\bf k}$  for the outer and inner layers, respectively. We assume that $\Gamma_{o,\bk}$ is impurity-dominated, and therefore isotropic and $T$-independent, while $\Gamma_{i,\bk}$ is highly anisotropic---either due to pseudogap physics or to order parameter fluctuations---and decreases with decreasing $T$.    
The $T$-dependent parameters that reproduce the oscillation frequencies and, qualitatively, the Fourier peak heights of Fig.~\ref{Fig2}(c) are given in Table~\ref{Table1}.

\begin{table}
\begin{tabular}{c|c|c}
& $T=4.2$~K &  $T=1.8$~K \\
\hline 
$M$  & 0.3 & 0.05 \\  
$\Delta_o$ & 0.1  & 0.15 \\
$\Delta_i$ & 0.0  & 0.1 \\
$\Gamma_{o,\bf k}$ & $0.020$   & $0.020$ \\
$\Gamma_{i,\bf k}$ & $0.005+0.005\eta_{\bf k}^2$  &  $0.005+0.002\eta_{\bf k}^2$ 
\end{tabular}
\caption{$T$-dependent order parameters and scattering rates used to simulate quantum oscillations at $T=4.2$~K and $T=1.8$~K.
Values are in units of the outer-layer nearest-neighbor hopping matrix element, $t_{o1} \sim 170$~meV and $\eta_{\bf k} = \cos k_x - \cos k_y$.  Tight-binding band parameters (given in the SI) are held constant. }
\label{Table1}
\end{table}

We calculated the density of states at the Fermi energy, $N_0(H)$, as a function of magnetic field using a numerical recursion method for large supercells. Results are shown in Fig.~\ref{Fig_theo}(a) for parameter sets corresponding to $T=4.2$~K and $T=1.8$~K.  The 4.2~K oscillations are dominated by a single frequency, while multiple frequencies are evident at 1.8~K.  Fourier transforming $N_0(H)$ with respect to $1/H$ reveals a strong low-frequency peak at $F\approx 600$~T at 4.2~K, along with harmonics at 1200~T and 1800~T. The peak  weakens and shifts to a higher frequency, $F\approx 800$~T, at 1.8~K.  Additional peaks, most notably at 1100~T and 2700~T, emerge at low $T$. These frequencies, which are extremely sensitive to model details, agree surprisingly well with the experiments.

The full reconstructed Fermi surfaces in the CDW state are too complicated to allow a semiclassical interpretation (see SI).  We  plot, instead, simplified Fermi surfaces in Fig.~\ref{Fig_theo}(c) and (d):  the bonding bands, which have primarily inner-layer character, are reconstructed by AFM order to form hole (blue) and electron (grey) pockets;  the primarily outer-layer antibonding (solid red) and nonbonding (solid green) Fermi surfaces are unchanged by AFM order, but are reconstructed by the CDW to form electron pockets (dashed red and green). These plots demonstrate that the main peak at $T$ = 4.2~K comes from the  hole pockets, whose areas match the peak frequencies.  It follows that the frequency shift with decreasing $T$ occurs because the hole pockets grow as $M$ decreases.  The peak-height reduction with decreasing $T$ comes from processes that disrupt semiclassical orbits along the hole pockets.  By carefully tracking the peak height for a range of parameter values, we can identify at least two key processes: quasiparticle scattering by the CDW between Fermi surface points that are separated by $\pm\bq_a$ and $\pm\bq_b$, and magnetic breakdown when $M$ is small. In Fig.~\ref{Fig_theo}(c), $q$ is too small to connect pairs of hole pockets, but may scatter quasiparticles onto the antibonding or nonbonding Fermi surfaces.  This is a weak process because the hole band is centered on a different layer than the other two.  However, CDW scattering grows rapidly as $M$ decreases and dominates when $M$ is  small enough that the hole pockets are connected by $q$ [Fig.~\ref{Fig_theo}(d)].  Then, the hole Fermi surfaces are reconstructed to form a diamond-shaped electron pocket and the main peak collapses rapidly.  This introduces a new oscillation frequency, $F\sim 2700$~T, that is clearly visible in Fig.~\ref{Fig_theo}(b) and provides an explanation for the high-frequency peak seen experimentally [Fig.~\ref{Fig2}(c)].  Other low-$T$ peaks, at 1100~T and 2100~T, are clearly tied to the emergence of inner-layer CDW order, but are difficult to ascribe to a single semiclassical orbit.

\begin{figure} [t]
\includegraphics[width=\columnwidth]{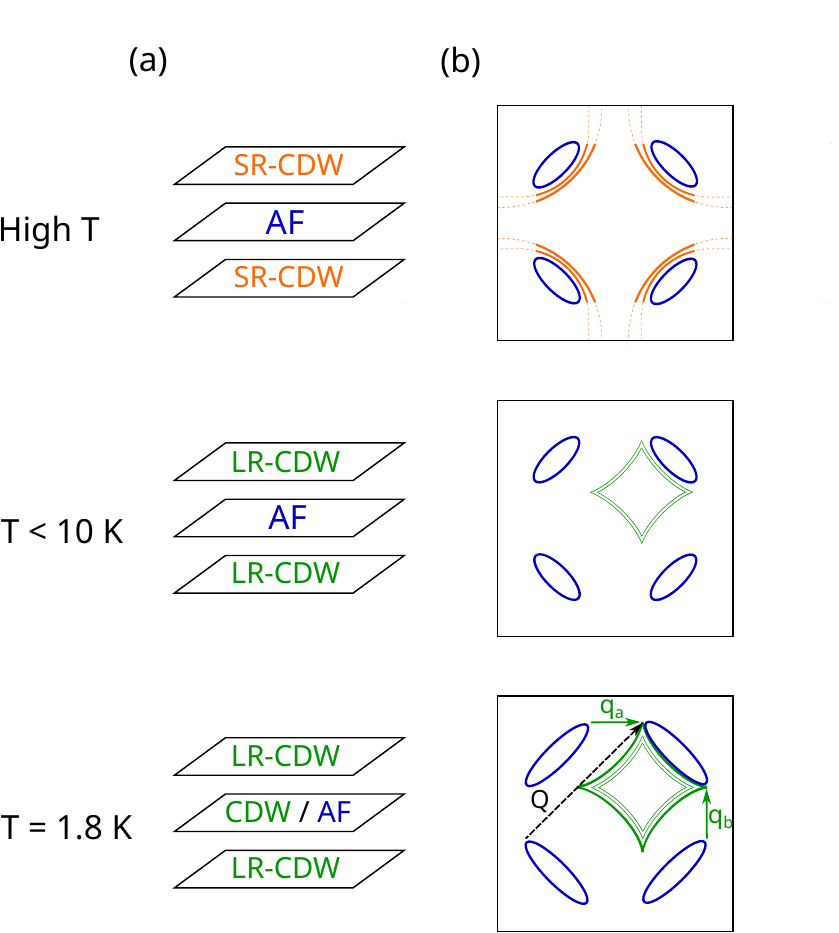}
\caption{\textbf{} a) Sketch of the trilayer structure of Hg1223 in different temperature range. SR(LR)-CDW stands for short-range (long-range) CDW and AF corresponds to antiferromagnetism. b) Sketch of the Fermi surface corresponding to the sequence of crossover as the temperature is reduced (see text).  ${\bf q}_a = (q, 0)$ and $\bq_b= (0, q)$ are the wavevector of the biaxial CDW. ${\bf Q}$ is the AFM wavevector.
}
\label{Fig_sketch}
\end{figure}
Our calculations support the scenario shown in Fig.~\ref{Fig_sketch}.  Above 10~K, inner-layer AFM order coexists with short-range outer-layer CDW order.   Below 10~K, the CDW is sufficiently long-range that the outer-layer Fermi surfaces are reconstructed and generate an electron-like Hall coefficient; however, the relatively large quasiparticle scattering rate in the outer layers damps the corresponding quantum oscillations. (In our simulations, the outer layers account for less than 10\% of oscillation amplitude.) Though centered on different layers, the AFM and CDW orders compete because the layers are coupled, and CDWs nucleate on the inner layer below 4.2~K at the expense of antiferromagnetism.  The low inner-layer quasiparticle scattering rate ensures that the crossover from AFM to CDW order qualitatively changes the quantum oscillations.  In particular, the reduction of the main peak height, its shift to higher frequency, and the appearance of the high frequency peak are intimately connected to the emergence of inner-layer charge order at the expense of antiferromagnetism.

A few remarkable implications of our work need to be discussed. Both experimental data and calculations point to a Fermi surface reconstruction in the inner plane from an AFM state (hole pockets) to a biaxial CDW (electron pocket), as discussed in ref.~\onlinecite{Atkinson:charge2015,Harrison18}. 
More specifically, AFM and CDW order parameters compete because they share the same Fermi-surface hotspots. The magnitude $q$ of the CDW wavevectors was chosen to explain the low-$T$ oscillation peak at $\sim 2800$~T. However, the CDW hotspots implied by this $q$ value are, within the resolution of our simulations, coincident with the AFM hotspots, i.e.\ points on the Fermi surface connected by the AFM wavevector ${\bf Q} = (\pi/a_0, \pi/a_0)$ (Fig.~\ref{Fig_sketch} bottom).  This coincidence is a key feature of models in which CDW order is mediated by critical spin fluctuations carrying momentum ${\bf Q}$.  As discussed above, we infer from our measurements that the AFM moment is very small, which implies that the AFM quantum critical point is nearby.  Our measurements are thus suggestive that charge order at this doping level is mediated by  critical spin fluctuations.

One possible counter-argument is that the superconducting and long-range CDW transition temperatures differ by an order of magnitude, contrary to early calculations suggesting they should be the same at the AFM quantum critical point \cite{Metlitski:2010instabilities,Efetov:pseudogap2013}; however, more recent work showed that the predicted degeneracy  breaks down when the band structure lacks particle-hole symmetry \cite{Wang:2018fragility}, as in Hg1223.  Another possible counter-argument is that CDW order persists in several cuprate families to doping levels far from the AFM quantum critical point \cite{Blanco-Canosa:2014,tam_charge_2022}.  It is unlikely that the AFM hotspot picture makes sense in these doping regimes, and other (noncritical) interactions are likely important.  The question posed by our work is whether the hotspot picture makes sense close to the AFM quantum critical point.
In this regard, it would be interesting to determine the evolution of the CDW wavevectors with doping because  the CDW hotspots should be pinned to the AFM Brillouin zone boundary wherever critical spin fluctuations are the dominant interaction.  

In conclusion, we have measured the evolution of the Fermi surface at low temperatures in Hg1223 samples with $p=0.10$-0.12, and demonstrated that there is a Fermi surface reconstruction at low temperature.  Through numerical simulations, we attribute this to competition between CDW and AFM order.  The CDW wavevectors required to reproduce the experimental oscillation spectrum suggest that the CDW and AFM phases share Fermi surface hotspots.  This naturally explains the competition between the two phases, but is also a hallmark of CDWs that are mediated by critical spin fluctuations.  

WAA acknowledges the support of the Natural Sciences and Engineering Research Council of Canada (NSERC).  This work was made possible by the facilities of the Shared Hierarchical 
Academic Research Computing Network (www.sharcnet.ca) and the Digital Research Alliance of Canada.
D.V. and C.P. acknowledges support from the EUR grant NanoX n\textsuperscript{o}ANR-17-EURE-0009 and from the ANR grant NEPTUN n\textsuperscript{o}ANR-19-CE30-0019-01. This work was supported by LNCMI-CNRS, members of the European Magnetic Field Laboratory (EMFL). CP wishes to thank the Institut Quantique in Sherbrooke for its hospitality.\\

\bibliographystyle{apsrev4-2} 

$^{\dagger}$ These authors contributed equally to this work.\\
$\parallel$ Present address: Laboratoire de Physique des Solides, 91405 Orsay, France\\
$^*$ billatkinson@trentu.ca\\
$^\ddagger$ cyril.proust@lncmi.cnrs.fr

\bibliography{Hg1223_FSR_biblio.bib}

\clearpage
\section*{Supplemental Information}

\section{Theory}
\subsection{Tight-Binding Model}

We adopt a tight-binding description of a single CuO$_2$ trilayer.  Previous tight-binding fits to ARPES data for optimal and overdoped trilayer \BiT \cite{Ideta10,Luo22} and underdoped five-layer Ba$_2$Ca$_4$Cu$_5$O$_{10}$(F,O)$_2$ \cite{Kunisada20} provide hints as to the band structure of underdoped Hg1223, but do not directly to determine its parameterization.  Informed by these experiments, we take what we believe is a plausible parameterization of the underdoped trilayer compound while acknowledging that it is undoubtedly imperfect.

\begin{table}[h]
\begin{tabular}{c|c||c|c}
\multicolumn{2}{c||}{outer layer} & \multicolumn{2}{c}{inner layer} \\
\hline
$t_{o0}$ & 1.10 & $t_{i0}$ & 0.53 \\
$t_{o1}$ & -1.00 & $t_{i1}$ & -0.70 \\
$t_{o2}$ & 0.40 & $t_{i2}$ & 0.2 \\
$t_{o3}$ & -0.05 & $t_{i3}$ & -0.035 \\
$t_{oo}$ & 0.10 & $t_{io}$ & 0.05 
\end{tabular}
\caption{Tight-binding parameters for the trilayer model.  The model parameters roughly follow tight-binding fits to ARPES measurements for overdoped \BiT; these suggest that $|t_{o1}| \sim 170$~meV  while $t_{i1}/t_{o1} \approx 0.7$ \cite{Luo22}.  Underdoped Hg1223 is not expected to have an identical band structure to overdoped \BiT, however, and some parameters (notably $t_{oo}$ and $t_{io}$) have been adjusted to provide better fits to QO experiments.}
\label{TableSI1}
\end{table}

The trilayer has two distinct layer types: two outer layers (denoted ``$o$'') and an inner layer (``$i$'').
The dispersions for the isolated layers, without interlayer coupling, are
\begin{equation} 
\epsilon_{\alpha{\bf k}} = t_{\alpha 0} +2t_{\alpha 1} (c_x+c_y) + 4t_{ \alpha2}c_xc_y + 2t_{\alpha 3} (c_{2x} + c_{2y}) 
\end{equation}
where $c_x \equiv \cos k_x$, $c_{2x} = \cos(2k_x)$, etc., and $\alpha = i, o$ refers to the layer type.  
The layers are coupled by two distinct forms of interlayer hopping:
\begin{eqnarray}
	t_{oo{\bf k}} &=& t_{oo} \\
	t_{io{\bf k}} &=& t_{io} \eta_{\bf k}^2
\end{eqnarray}
with $\eta_{\bf k} = c_x - c_y$.  The interlayer coupling in overdoped \BiT is reported to have a complicated ${\bf k}$-dependence \cite{Luo22}, and here we have simplified the model by taking a coupling $t_{oo{\bf k}}$ between outer layers that is ${\bf k}$-independent.  While this is likely an oversimplification, the outer layers are of secondary importance for quantum oscillation experiments because of their relatively high disorder levels.
The coupling $t_{io\bf k}$ between the inner and outer layers has a qualitatively similar structure to that reported in Ref.~\cite{Luo22}, but has a magnitude that is 60\% as large.  Importantly, $t_{io{\bf k}}$ vanishes along the Brillouin zone diagonals due to the factor $\eta_{\bf k}^2$, which is the result of a quantum filtering effect \cite{Pavarini01}.  Tight-binding parameters are given in Table~\ref{TableSI1}.  

\begin{figure}
\includegraphics[width=\columnwidth]{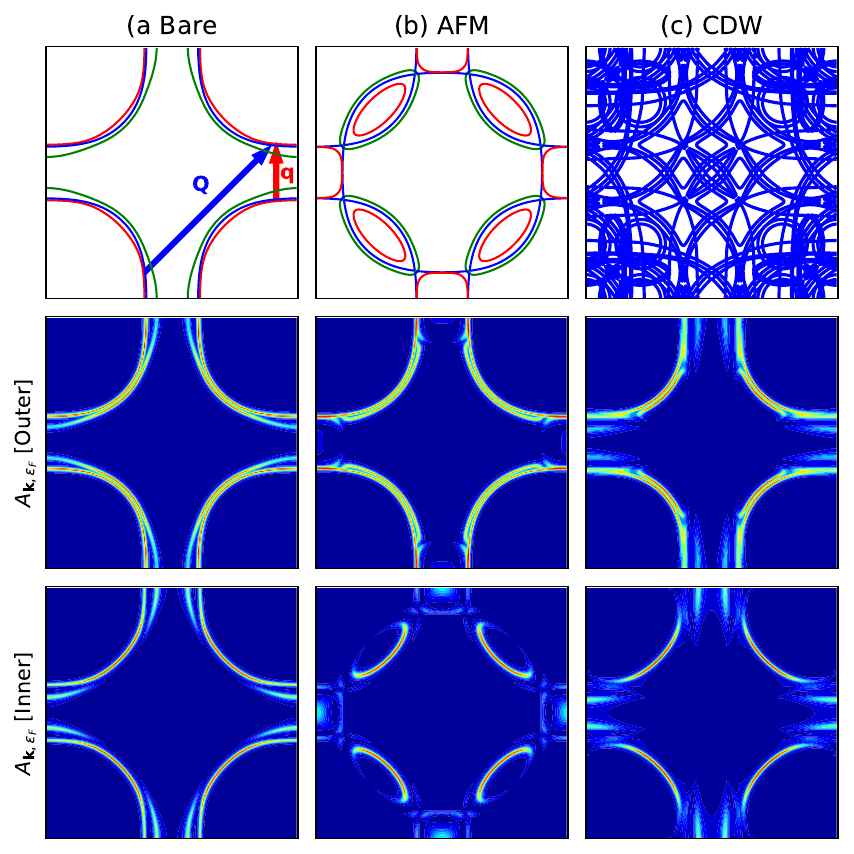}
\caption{Fermi surface structure for the model parameters in Table~\ref{TableSI1}.  Column (a) shows the bonding (green), nonbonding (blue), and antibonding (red) Fermi surfaces (top row).  The outer-layer spectral function (middle row), and inner-layer spectral function (bottom row) show the weights of each band in the different layers.  The CDW wavevector ${\bf q} = (0, 2\pi/4.5a_0)$ and the AFM wavevector $\QQ = (\pi/a_0, \pi/a_0)$ are also shown.    Column (b) shows the Fermi surfaces and spectral functions for the AFM case with $M=0.3$.  Column (c) shows the main Fermi surfaces and leading replicas for the cdw case with $\Delta_o = \Delta_i = 0.1$  The hole densities in the different layers are $p_i = 0.065$ and $p_o = 0.12$, with the overall hole density $p=(2p_o + p_i)/3 = 0.10$.}
\label{Fig:BandsSI}
\end{figure}

The trilayer Hamiltonian is $\hat H = \sum_\bk \Psi^\dagger_\bk  {\bf H}_\bk \Psi_\bk$, with $\Psi^\dagger_\bk = (c_{1\bk}^\dagger, c_{2\bk}^\dagger, c_{3\bk}^\dagger)$ and 
\begin{equation}
{\bf H}_\bk = \left[ \begin{array}{ccc}
\epsilon_{o\bk} & t_{io\bk} & t_{oo\bk} \\
t_{io\bk} & \epsilon_{i\bk} & t_{io\bk} \\
t_{oo\bk} & t_{io\bk} & \epsilon_{o\bk} \end{array}\right]
\label{eq:Hk}
\end{equation}
This Hamiltonian can be diagonalized easily and gives a bonding band, an antibonding band, and a nonbonding band.
\begin{eqnarray}
E_{{\bf k} \pm} &=& \frac{\epsilon_{o{\bf k}} + t_{oo{\bf k}} + \epsilon_{i{\bf k} }}{2}
\pm \left[ \left(\frac{\epsilon_{o{\bf k}} + t_{oo{\bf k}} - \epsilon_{i{\bf k} }}{2}\right )^2 + 2t_{io}^2\right ]^{1/2} 
\label{eq:Epm} \\
E_{{\bf k} 0} &=& \epsilon_{o{\bf k}} - t_{oo {\bf k}}
\label{eq:E0}
\end{eqnarray}
Fermi surfaces for these bands are plotted in Fig.~\ref{Fig:BandsSI}(a), along with the spectral functions in both the outer and inner layers.  The spectral functions show the projections of the different bands on the different layer types.  From this figure, we see that the nonbonding band has purely outer-layer character.  The bonding ($E_{\bk-}$) and antibonding ($E_{\bk+}$) bands have mixed character that depends on $\bk$.  Near the Brillouin zone diagonals (the ``nodal regions'') the bonding band has purely inner-layer character, while the antibonding band is purely outer-layer.   Near the Brillouin zone boundary (the ``antinodal regions''), on the other hand, the two bands have  mixed character, so that the spectral weights in the two layers are comparable for each band.

\subsubsection{Quasiparticle Scattering}
The spectral functions shown in Fig.~\ref{Fig:BandsSI} are made assuming that the quasiparticle lifetime is isotropic.  In our full calculations of the quantum oscillation spectrum, the scattering rate is taken to be anisotropic via an imaginary layer-dependent self-energy $i\Gamma_{\alpha {\bf k}}$ where 
\begin{eqnarray}
\Gamma_{\alpha {\bf k}} = \Gamma_{\alpha 0 } + \Gamma_{\alpha 1 } \eta_{\bf k}^2,
\end{eqnarray}
with $\alpha=i, o$ indicating layer type and $\eta_\bk = \cos k_x - \cos k_y$.  The isotropic component, $\Gamma_{\alpha 0}$, is attributed to impurity scattering.  Because the outer layers are adjacent to  dopant layers with random disorder, $\Gamma_{o0}$ is much larger than the isotropic component $\Gamma_{i0}$ for the inner layers.  However, the inner layers  have a  low hole concentration, and should be strongly affected by the pseudogap.  We capture one aspect of the pseudogap physics, namely the short quasiparticle lifetimes in the antinodal region, by the anisotropic scattering rates $\Gamma_{\alpha 1}$.  For our calculations, this term is most important for the inner layer, as that is the dominant source for our quantum oscillations.  Our model does not include the reduced quasiparticle spectral weight at the antinodes, which would require inclusion of strong correlations.

\subsubsection{Antiferromagnetic Order}
Antiferromagnetic order is allowed on the inner layer only.  We define an order parameter $M$, which in a weak-coupling picture is $M \sim U m$, where $U$ is the Hubbard $U$ and $m$ is the local magnetic moment in dimensionless units. The AFM order scatters quasiparticles through the wavevector $\QQ = (\pi/a_0, \pi/a_0)$.

In our simulations, $M\lesssim 0.3$ in units of $t_{o1}$, which corresponds to $\sim 50$~meV assuming $t_{o1} = 170$~meV.  The Fermi surfaces and spectral functions for $M=0.3$ are shown in Fig.~\ref{Fig:BandsSI}(b).  Although the Fermi surface contours suggest that all Fermi surfaces are reconstructed by the AFM order, the spectral functions reveal that the reconstruction is substantial only in the inner layer, while the energy scale for the AFM gap  in the outer layers induced by the proximity effect is small.

\subsubsection{Charge Density Wave Order}
The CDW order can be included either as site-centered ($s$-wave) shifts of the site energy, or bond-centered shifts of the hopping amplitude. In this work, we model bond-centered CDWs with a $d$-wave form factor.  The CDW is biaxial and therefore has Fourier components at ${\bf q} = (\pm q, 0)$ and $(0, \pm q)$.  This is represented by shifts in the hopping matrix elements as follows:  the hopping matrix element from site $\ell$ to site $\ell+x$ (that is, a hop by a single lattice constant in the $x$-direction) is 
\begin{equation}
t_{\alpha \ell+x\, \ell} = t_{\alpha 1} + \Delta_\mathrm{\alpha}[ \cos(q x_{\ell+0.5x}) + \cos(q y_{\ell})],
\label{eq:tx}
\end{equation}
while hopping from $\ell$ to $\ell+y$ is
\begin{equation}
t_{\alpha \ell+y\, \ell} =  t_{\alpha 1} - \Delta_\mathrm{\alpha}[ \cos(q x_{\ell}) + \cos(q y_{\ell+0.5y})],  
\label{eq:ty}
\end{equation}
with $\alpha$ the layer index.

For practical calculations, the CDW must be commensurate with an enlarged supercell containing an integer number of primitive unit cells.  This restriction makes it difficult to tune the CDW periodicity.  The situation is made more complicated in calculations of quantum oscillations because the supercell must also contain an integer number of magnetic flux quanta for \textit{every} choice of field.  We have considered several $q$-values for which both criteria could be satisfied and have found that $q = 2\pi/4.5a_0$ reproduces many of the features of the experimental QO spectrum.  In particular, the emergence of a low-$T$ peak at 2700~T requires an electron pocket that has an area that is much larger than is normally found in the CDW phase of cuprates.  We have only succeeded in obtaining this for a CDW wavevector that nests the antinodal regions of the Fermi surface, as illustrated in Fig.~\ref{Fig:BandsSI}(a).

A biaxial period-$4.5a_0$ CDW requires that the minimal supercell contain $9\times 9$ primitive unit cells, for a total of 243 bands.  While this physics is captured by the real-space approach used to calculate the QO spectrum (described below), we present in Fig.~\ref{Fig:BandsSI}(c) the results of a simplified calculation that captures the essential physics of the CDW phase.  The CDW Hamiltonian has the form 
\begin{equation}
H_\mathrm{CDW} = \sum_\bk \left [ 
\Psi^\dagger_\bk {\bf H}_\bk \Psi_\bk 
+ \sum_j \left ( \Psi^\dagger_{\bk+\bq_j} {\bf \Delta}_{\bk+\frac{\bq_j}{2}} \Psi_\bk + h.c. \right ) \right],
\label{eq:Hkq}
\end{equation}
with ${\bf H}_\bk$ given by Eq.~(\ref{eq:Hk}), where ${\bq}_j$ refers to the four possible CDW wavevectors, and where the matrix of CDW order parameters is ${\bf \Delta}_\bk = \mathrm{diag}(\Delta_{o\bk}, \Delta_{i\bk}, \Delta_{o\bk})$, with
\begin{equation}
\Delta_{\alpha \bk} = \Delta_\alpha (\cos k_x - \cos k_y).
\end{equation}
To second order in ${\bf \Delta}_\bk$, the Hamiltonian generates the Fermi surfaces and spectral functions shown in Fig.~\ref{Fig:BandsSI}(c). We emphasize that this is only a small subset of the total set of Fermi surfaces that would be produced by an exact solution of Eq.~(\ref{eq:Hkq}).  The key point of  Fig.~\ref{Fig:BandsSI}(c) is that the complex Fermi surface structure that emerges from this calculation reveals nothing about the structure of the spectral functions shown in the figure.  Physical properties, such as the density of states, are more closely linked to the spectral functions than the Fermi surfaces.  These can be understood with simplified band structure models, as descrbed in the main text.

\subsection{Real-space formulation}
\subsubsection{Model in real space}
The density of states at a site $\ell$ in layer $\alpha$ can be obtained from the Green's function via
\begin{equation}
N(\alpha \ell,\omega) = -\frac 1\pi \mbox{Im }\langle \alpha \ell | {\bf G}(\omega) | \alpha \ell \rangle.
\label{eq:N}
\end{equation}
where, using a general matrix notation,
\begin{equation}
{\bf G}^{-1}(\omega) = \omega \mathbf{1} - \mathbf{ H }+ i \mathbf{ \Gamma}.
\label{eq:Ginv}
\end{equation}
Here ${\bf H}$ is the Hermitian matrix form of the Hamiltonian and $i\mathbf{\Gamma}$ is the non-Hermitian matrix representing the anisotropic quasiparticle scattering.  The density of states is most easily evaluated when matrices are represented in the basis of sites ($i,j$) and layers ($\alpha, \beta$).

Then, the zero-field Hamiltonian has the simple block form 
\begin{equation}
\mathbf{H} = \left [ \begin{array}{ccc} 
{\bf H}_{oo} & {\bf T}_{io} & {\bf T}_{oo} \\
{\bf T}_{io} & {\bf H}_{ii} & {\bf T}_{io} \\
{\bf T}_{oo} & {\bf T}_{io} & {\bf H}_{oo} \end{array} \right ].
\end{equation}
We can obtain the matrix elements from the $\bk$-space dispersions via the inverse Fourier transform $X_{ij} = N^{-1} \sum_{\bf k} X_{\bf k} e^{i {\bf k}\cdot ({\bf r}_i - {\bf r}_j)}$.  The bare intralayer matrix elements obtained from $\epsilon_{\alpha\bk}$ are
\begin{equation}
[{\bf H}_{\alpha\alpha}^0]_{\ell m} = 
t_{\alpha 0} \delta_{\ell, m} + t_{\alpha 1} \delta_{\langle \ell, m\rangle} 
+ t_{\alpha 2} \delta_{\langle \langle \ell, m\rangle \rangle} 
+ t_{\alpha 3} \delta_{\langle \langle \langle \ell, m\rangle \rangle \rangle},
\end{equation}
where we use a generalized Kronecker-delta notation:  
$\delta_{\ell, m}=1$ when $\ell=m$ and is zero otherwise;  delta-functions with $n$-pairs of brackets in their subscript equal one when $\ell$ and $m$ are $n$th-nearest neighbors, and zero otherwise.  
To these bare matrix elements, we add contributions from the antiferromagnetism and CDW:
\begin{eqnarray}
[{\bf H}_{\alpha\alpha}]_{\ell m} &=& [{\bf H}_{\alpha\alpha}^0]_{\ell m} + M e^{i\QQ\cdot \br_\ell} \delta_{\ell, m} \delta_{\alpha, b} \nonumber \\
&&+ \Delta_\mathrm{\alpha}[ \cos(q x_{m+0.5x}) + \cos(q y_{m})] \delta_{\ell, m+x}
 \nonumber \\
&&- \Delta_\mathrm{\alpha}[ \cos(q x_{m}) + \cos(q y_{m+0.5y})] \delta_{\ell, m+y}.
\label{eq:Hr}
\end{eqnarray}
where ${\bf Q} = (\pi/a_0,\pi/a_0)$ is the AFM wavevector.

Similarly, the interlayer matrix elements are obtained from the Fourier transforms of  $t_{oo\bk}$ and $t_{io\bk}$, 
\begin{eqnarray}
{}[{\bf T}_{oo}]_{\ell m} &=& t_{oo}\delta_{\ell, m} \\
{}[{\bf T}_{io}]_{\ell m} &=& t_{io}\left ( \delta_{\ell, m} 
- \frac 12 \delta_{\langle \ell, m \rangle}
+ \frac 14 \delta_{\langle\langle\langle \ell, m \rangle\rangle\rangle}
\right ),
\end{eqnarray} 
while the quasiparticle scattering matrix elements are 
\begin{equation}
[ {\bf \Gamma}_{\alpha\alpha} ]_{\ell m}  = (\Gamma_{\alpha 0} + \Gamma_{\alpha 1}) \delta_{\ell, m} 
- \frac 12 \Gamma_{\alpha 1} \delta_{\langle \ell, m \rangle}
+ \frac 14 \Gamma_{\alpha 1} \delta_{\langle\langle\langle \ell, m \rangle\rangle\rangle}.
\end{equation}

\subsubsection{Peierls substitution}
\begin{figure}
\includegraphics[width=\columnwidth]{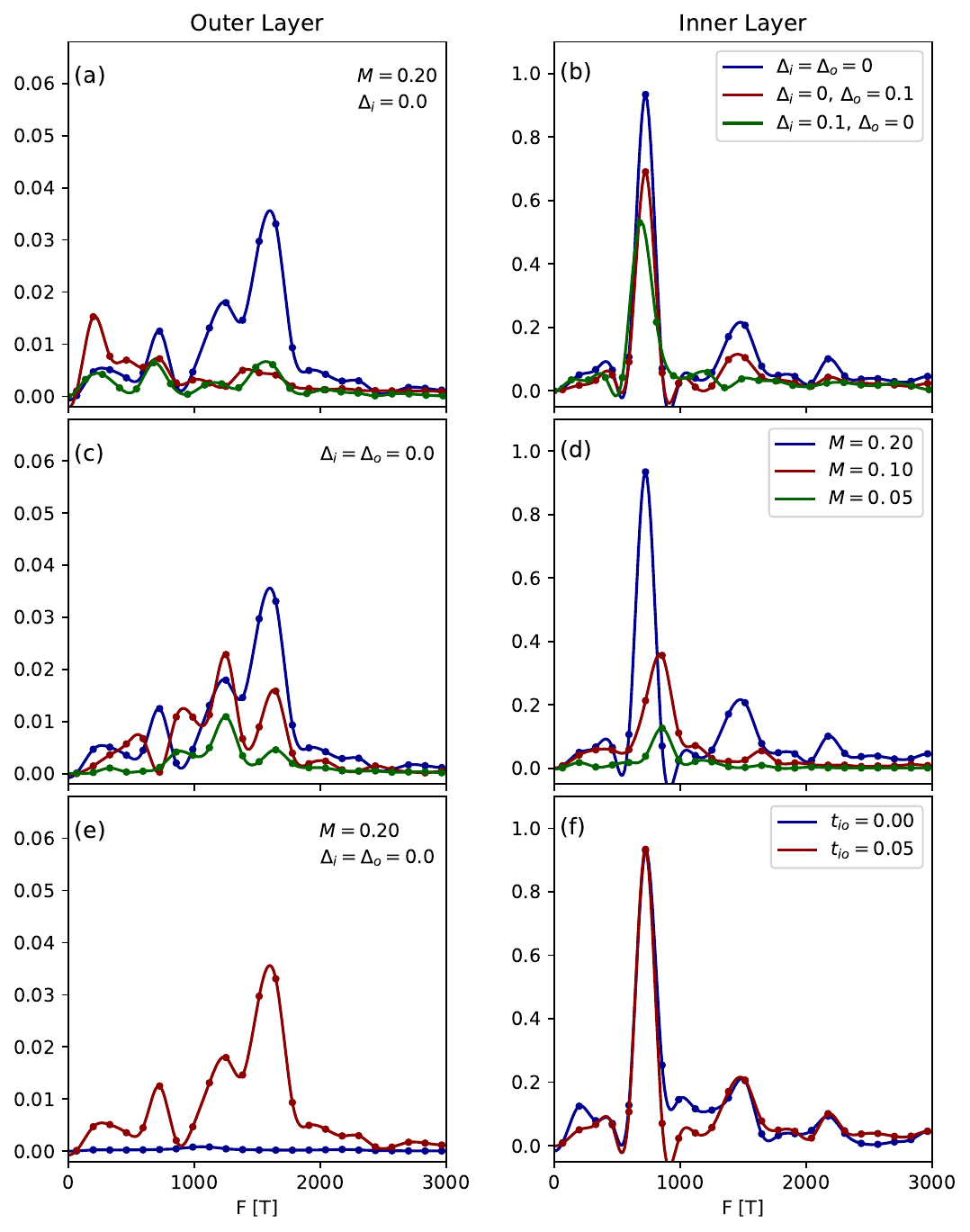}
\caption{Effect of charge order, antiferromagnetism, and interlayer coupling on the main peak of the Fourier-transformed density of states.  (a), (b)  Evolution  with  CDW order in either the outer or inner layer for fixed $M=0.20$.  (c), (d) Evolution with $M$ for $\Delta_i = \Delta_o = 0.0$.  (e), (f) Effect of switching off the hopping between inner and outer layers.  In all figures, $\Gamma_{i\bk} = 0.005+0.005\eta_\bk^2$ and $\Gamma_{o\bk} = 0.02$.  The left column shows results for a single outer layer, while the right column shows results for the inner layer.}
\label{fig:pk_SI}
\end{figure}
An orbital magnetic field can be added to the Hamiltonian, Eq.~(\ref{eq:Hr}) via a Peierls substitution, 
\begin{eqnarray}
\tilde c_{\alpha \ell} &=&  c_{\alpha \ell} \exp\left (i\frac{e}{\hbar}\int_{\br_0}^{\br_\ell} \bA(\br')\cdot d\br'\right ) \\
{} [{\bf \tilde H}_{\alpha \beta}]_{\ell m} &=& [{\bf H}_{\alpha \beta}]_{\ell m}
\exp\left (i\frac{e}{\hbar}\int_{\br_\ell}^{\br_m} \bA(\br')\cdot d\br'\right )
\end{eqnarray}
where $c_{\alpha i}$ is the electron annihilation operator at lattice site $i$ in layer $\alpha$ and $\br_0$ is a reference point in the lattice.
A uniform magnetic field perpendicular to the layers, ${\bf B} = B_0 {\bf \hat z}$, can be obtained from the magnetic vector potential 
\begin{equation}
	\bA = B_0 x {\bf \hat y}
\end{equation}
such that
\begin{equation}
	 [{\bf \tilde H}_{\alpha\beta}]_{\ell m} = [{\bf H}_{\alpha\beta}]_{\ell m} \exp\left[ i\frac{eB_0}{\hbar} \frac{x_m + x_\ell}{2} (y_\ell-y_m) 
	\right ].
\end{equation}
We define a magnetic supercell of dimensions $L_x \times L_y$ such that it contains an integer number of magnetic flux quanta,
\begin{equation}
	\frac{eB_0}{\hbar}L_xL_y = 2n\pi \quad \mbox{or} \quad \Phi = n\Phi_0,
\end{equation}
with $\Phi_0 = 2\pi \hbar/ e$ and $\Phi = B_0L_xL_y$.   Thus
\begin{equation}
[{\bf \tilde H}_{\alpha \beta}]_{\ell m} = [{\bf H}_{\alpha \beta}]_{\ell m} \exp\left \{ i\alpha \left[ \frac{x_\ell + x_m}{2} + \frac{X_L + X_M}{2} 
\right ]\Delta y 
\right \}
\end{equation}
where $\alpha = 2n\pi/L_xL_y$, $\Delta y = y_\ell - y_m + Y$,  $Y = Y_L-Y_M$, and $X = X_L - X_M$.  
In this expression, and from now on, lower-case symbols (e.g.\ $x_\ell$, $y_m$) refer to positions within the supercell and $X_L$ and $Y_M$ refer to coordinates of the supercells. 

We can express this in terms of magnetic Bloch states,
\begin{equation}
	\tilde c_{{\bf K}m} = \frac{1}{\sqrt{n_k}}\sum_M c_{Mm} e^{i\alpha y_m X_M} e^{-i{\bf K\cdot R}_M},
\end{equation}
with
\begin{equation}
{}	[{\bf \tilde H}_{\alpha\beta}({\bf K})]_{\ell m} =  [ {\bf H}_{\alpha\beta}]_{\ell m} e^{i\alpha \left[ \frac{x_\ell + x_m}{2}\Delta y - \frac{y_\ell+y_m}{2} X + \frac{XY}2 \right ]}
	e^{-i{\bf K}\cdot{\bf R}}
\end{equation}
In Hg1223, the lattice constant is $a_0 \approx 3.85$~\AA, and to obtain a field $B_0 =50$~T, we need $L_xLy = \Phi_0/B_0 \approx 500$ if the supercell is to contain a single flux quantum.  This can be achieved with modest supercell sizes; however, the requirement that both $L_x$ and $L_y$ be commensurate with the CDW period increases the required size of $L_x$ and $L_y$ by an order of magnitude or more.  This generally rules out standard matrix diagonalization techniques as a route to obtaining the density of states.  

We evaluate the density of states directly using a generalization of the recursion method first introduced by Haydock \cite{Haydock80}.  Haydock's approach assumes that the inverse Green's function has the form 
\begin{equation}
{\bf G}^{-1} = (\omega + i\Gamma) \mathbf{1} - \mathbf{H},
\end{equation}
where $\mathbf{H}$ is a Hermitian matrix.  In the current work, $\mathbf \Gamma$ is different in the inner and outer layers and ${\bf G}^{-1}$ is instead given by Eq.~(\ref{eq:Ginv}).  In this case, we can generalize Haydock's method to employ a two-sided Lanczos algorithm to obtain the projection of the Green's function onto orbital $|\alpha \ell\rangle$, as in Eq.~(\ref{eq:N}).

\subsection{Additional Results: Reduction of the main peak height}
To determine the reason for the main peak height reduction in our simulations, we compare in Fig.~\ref{fig:pk_SI} the  Fourier transformed density of states for different parameter sets.   Figure~\ref{fig:pk_SI}(a) and (b) show the effect of CDW order on the main peak for fixed $M=0.20$.  Results are shown for CDW order in the outer layers only ($\Delta_o =0.1$, $\Delta_i = 0$) and the inner layer only ($\Delta_o = 0$, $\Delta_i = 0.1$).  Until $M$ is small enough that the CDW nests the tips of the AFM hole pockets, CDW order has a modest effect on the peak height.  Figures~\ref{fig:pk_SI}(c) and (d)  show that the peak height drops rapidly as $M$ is decreased, even when there is no CDW.  This implies that magnetic breakdown plays a key role.  The two obvious magnetic breakdown processes involve tunneling from the AFM hole pocket to the AFM electron pocket, and tunneling to the antibonding or nonbonding Fermi surfaces.  The latter of these depends on the coupling between inner and outer layers, and in Fig.~\ref{fig:pk_SI}(e) and (f), we show that setting $t_{io} = 0$ has negligible effect on the peak height, which suggests that tunneling between the bands is less important than tunneling between the hole and electron pockets.

\section{Experiment}
\begin{figure} [h] 
\centering
\includegraphics[width=0.4\textwidth]{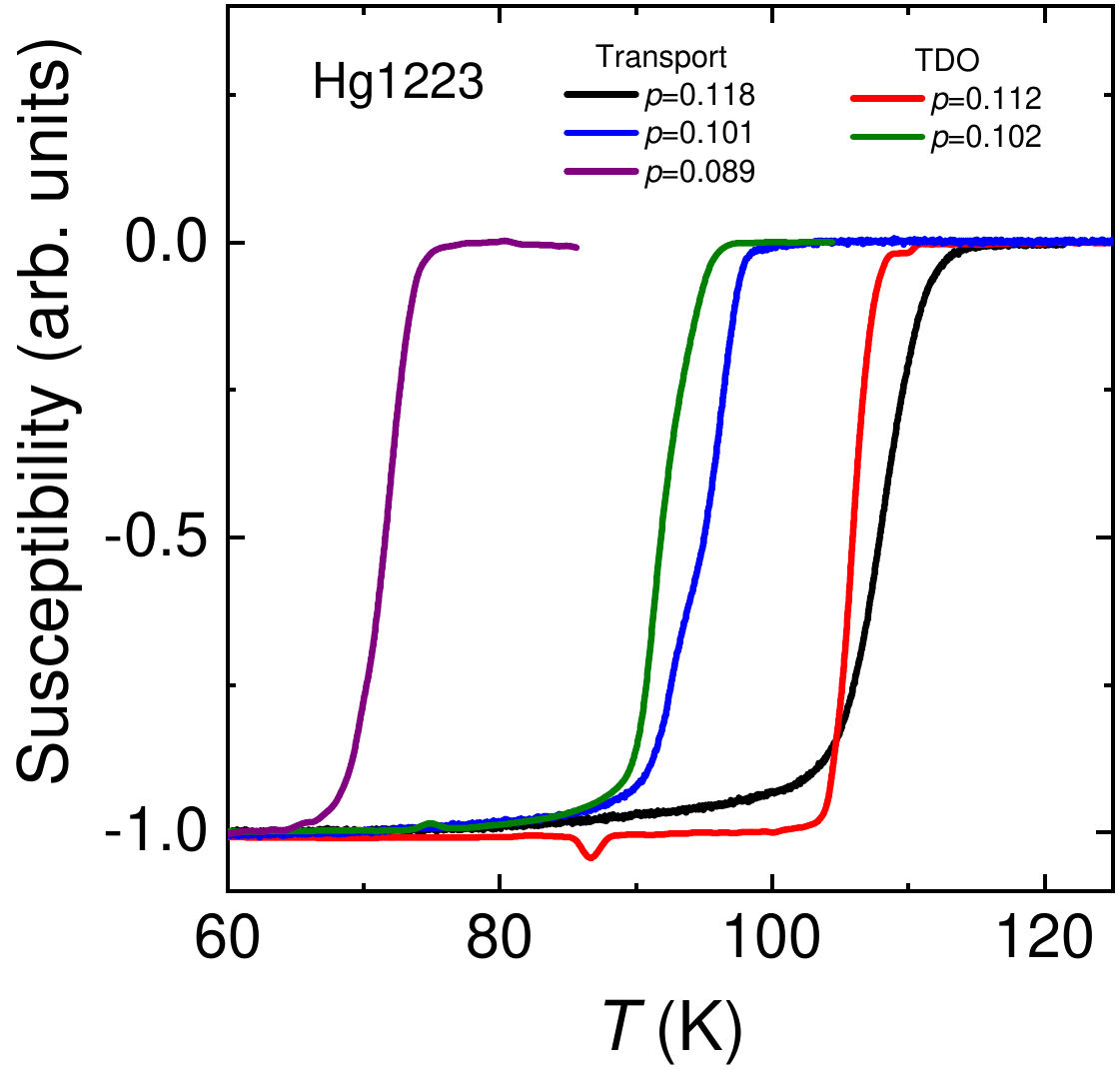}
\caption{Magnetic susceptibility measurements using SQUID of vacuum-annealed crystals of \HgT at different doping levels, as indicated. 
}
\label{FigS1}
\end{figure}
\subsection{List of samples and magnetization measurements}
\begin{table}[h]
\centering
 \begin{tabular}{|c|c|c|c|c|c|}
  \hline
   sample & $T_c$  & $p$ & measurements \\
  \hline
  Hg1223 & 114~K & 11.8~\% & transport  \\
  \hline
  Hg1223 & 108~K & 11.2~\% & TDO \\
  \hline
  Hg1223 & 95~K & 10.1~\% & transport \\
  \hline
  Hg1223 & 96~K & 10.2~\% & TDO \\
  \hline
  Hg1223 & 78~K & 8.9~\% & transport \\
  \hline
\end{tabular}
\caption{\label{mc} $T_c$, hole doping and probe measurement for the five Hg1223 samples measured in this study.  } 
\end{table}  
Single crystals of the trilayer cuprate \HgT have been synthesized using a self-flux growth technique as described in ref.~\onlinecite{Loret17}. Using adequate heat treatment, Hg1223 can be largely underdoped and its doping level controlled. The doping $p$ has been deduced from the empirical relation $1 - T_c / T_{c,max} = 82.6(p - 0.16)^2$, where $T_c$ is the onset superconducting transition measured by SQUID (see Supplementary Fig.~\ref{FigS1}) and $T_{c,max}$ = 133~K.

\subsection{Hall effect in Hg1223}
\begin{figure*} [t]
\includegraphics[width=1\textwidth]{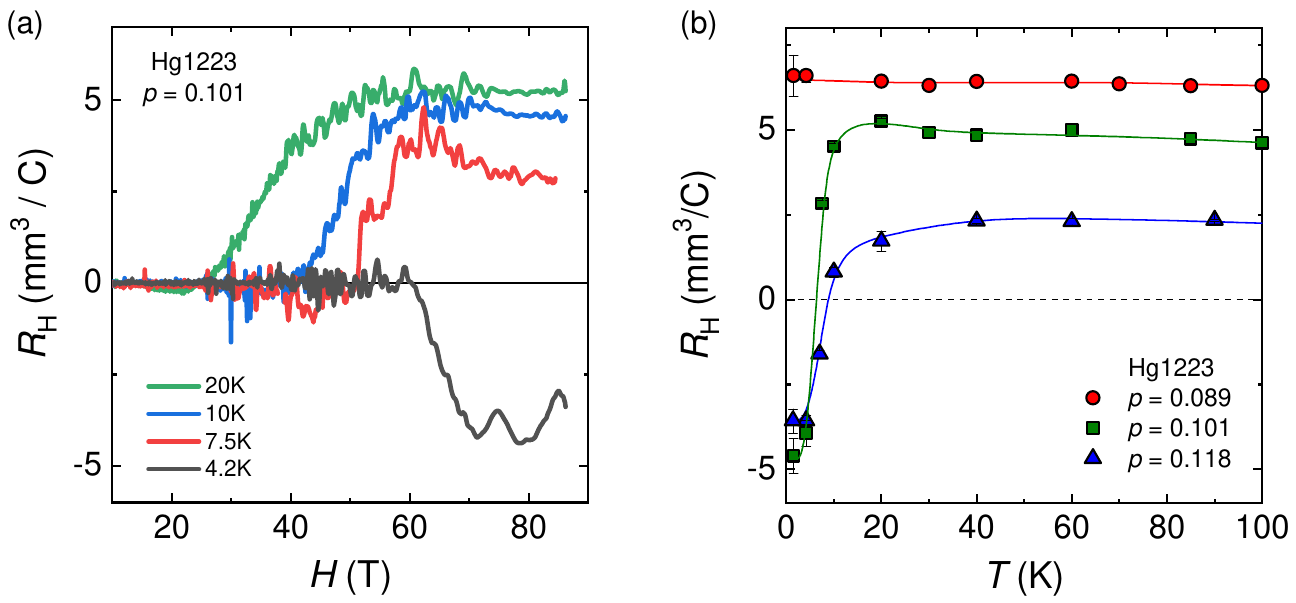}
\caption{\textbf{} a) Field dependence of the Hall coefficient $R_H$ in Hg1223 ($p$ = 0.101) at various fixed temperatures, as indicated.
b) Temperature dependence of the normal-state Hall coefficient $R_H$, measured at high fields, in Hg1223 $p$ = 0.09 (red circles from \cite{Oliviero22}, $p$ = 0.101 (green squares, this work)  and $p$ = 0.181 (blue triangles, this work). At $p$ = 0.118 and $p$ = 0.101, $R_H$ changes sign abruptly below $T$ = 10~K  while it remains positive down to the lowest temperature for $p$ = 0.09.
}
\label{SI_Hall}
\end{figure*}
The Hall effect was measured in Hg1223 at doping level $p$ = 0.118 and $p$ = 0.101. Data for the sample $p$ = 0.09 is from ref.~\onlinecite{Oliviero22}. Typical sample dimensions  are 700*400*90~$\mu$m$^3$. Gold contacts were sputtered onto the surface of the sample before a heat treatment leading to contact resistances of a few ohms at room temperature and below 1~$\Omega$ at low temperature. The magnetic field $H$ was applied along the $c$-axis of the tetragonal structure, perpendicular to the CuO$_2$ planes in both polarities of the field.
The measurements were performed up to 86 T in a dual coil magnet using a conventional 4-point configuration with a current excitation of $\approx$ 5~mA at a frequency of $\approx$ 60 kHz. A high-speed acquisition system was used to digitize the reference signal (current) and the voltage drop across the sample at a frequency of 500 kHz. The data was post-analyzed with a software to perform the phase comparison.\\
Fig.~\ref{SI_Hall}a shows the Hall coefficient plotted as a function of magnetic field H for Hg1223 $p$ = 0.101 at different temperatures from $T$ = 20~K down to $T$ = 4.2~K. While $R_H$ is positive above $T$ = 20 K, it become negative at $T$ = 4.2~K.  The temperature dependence of the normal-state Hall coefficient measured at the highest fields between $T$ = 1.5~K and 100~K is shown in Fig.~\ref{SI_Hall}b for Hg1223 at three doping levels, $p$ = 0.101 and $p$ = 0.118 (this work) and $p \approx$~0.09 \cite{Oliviero22}. For $p$ = 0.101 and $p$ = 0.118, there is a sudden sign change of the Hall coefficient below $T \approx$~ 10~K that is not observed at lower doping ($p$ = 0.09) down to the lowest temperature.

\subsection{Quantum oscillations in Hg1223}
Quantum oscillations have been measured using a contactless tunnel diode oscillator-based technique \cite{Coffey00} in two samples of Hg1223 at doping level $p$ = 0.112 and $p$ = 0.102. Typical sample dimensions are 500*500*100~$\mu$m$^3$. The experimental setup consists of a LC-tank circuit powered by a tunnelling diode oscillator biased in the negative resistance region of the current-voltage characteristic. The sample is placed in a compensated 8-shape coil (diameter and length of the coil are adapted for each sample to optimize the filling factor). The fundamental resonant frequency $f_0$ of the whole circuit is about 25 MHz. The RF signal is amplified and demodulated down to a frequency of about 1~MHz using a heterodyne circuit. A high-speed acquisition system is used to digitize the signal. The data are post-analysed using a software to extract the field dependence of the resonance frequency $f_{TDO}$, which is sensitive to the resistivity through the change in skin depth.
%
\begin{figure*} [t]
\includegraphics[width=1\textwidth]{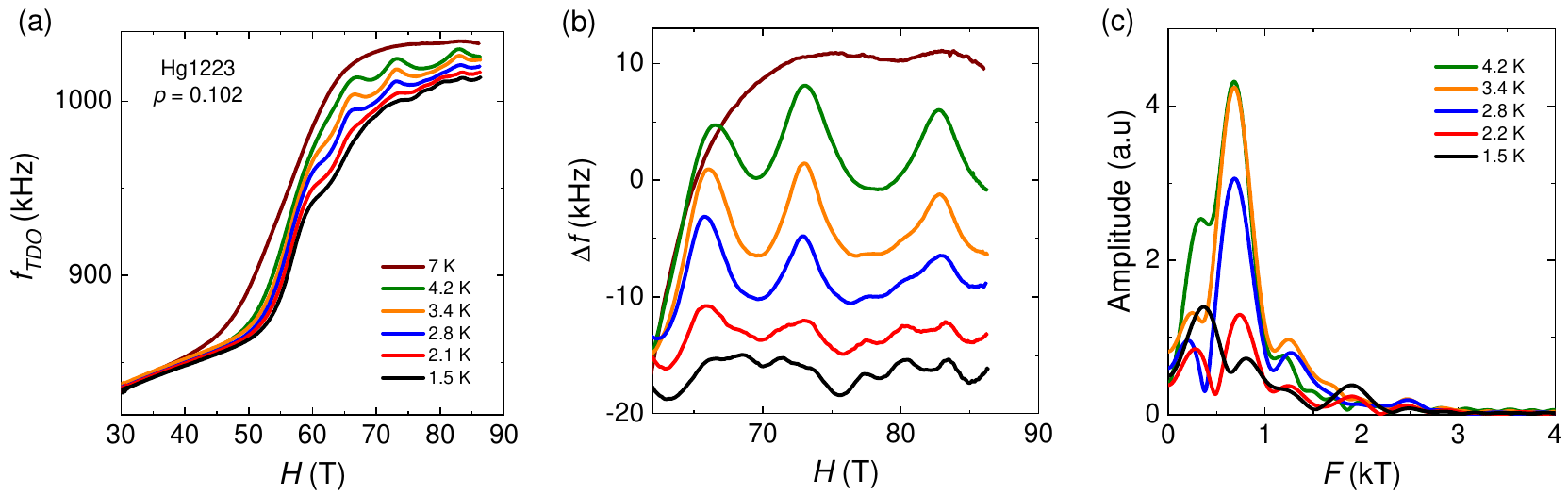}
\caption{\textbf{} a)  Field dependence of the TDO frequency in  Hg1223 ($p$ = 0.102) at various fixed temperatures, as indicated. 
b) Oscillatory part of the TDO signal after removing a smooth background (spline) from the data shown in panel a).
c) Discrete Fourier analysis of the oscillatory part of the TDO signal shown in panel b).  
}
\label{QO10}
\end{figure*}
\begin{figure*} [t]
\includegraphics[width=1\textwidth]{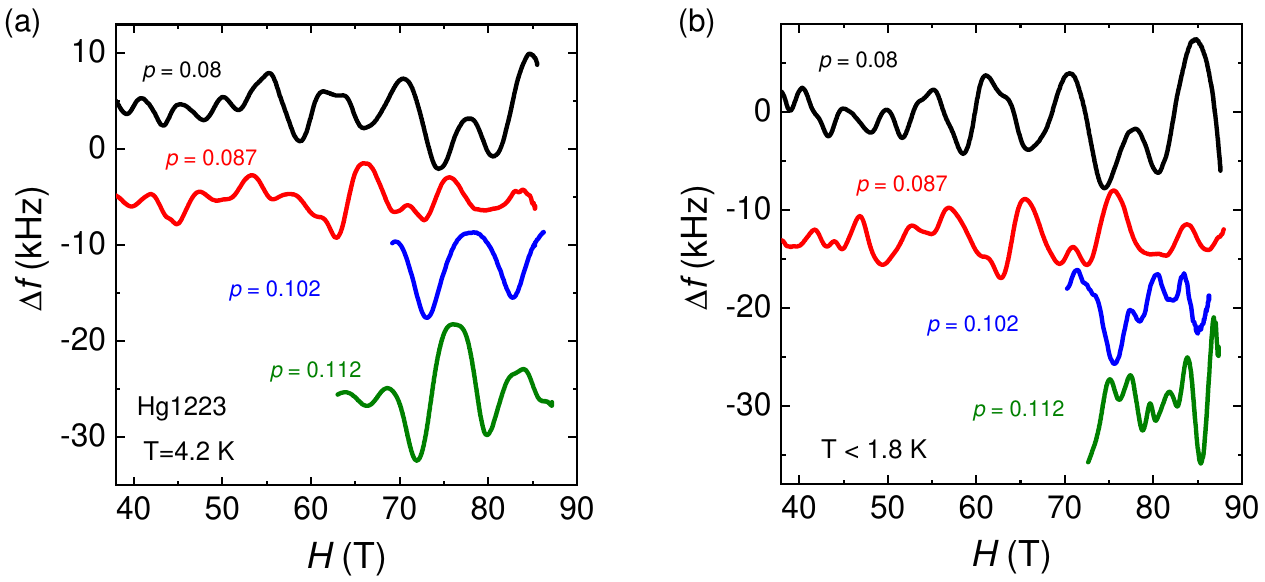}
\caption{\textbf{} Quantum oscillation spectra in Hg1223 at various doping level, as indicated, measured at
a) $T$ = 4.2 K and b) $T <$ 1.8~K. Data at $p$~=~0.08 and $p$~=~0.087 are taken from ref.\cite{Oliviero22}.  
}
\label{QOall}
\end{figure*}
%
In Supplementary Fig.~\ref{QO10}, we show the evolution of the quantum oscillation spectrum across the transition in Hg1223 $p$ = 0.102. Fig.~\ref{QO10}a shows the raw data as a function of magnetic field at different temperatures. A smooth background subtraction leads to the oscillatory part of the signal depicted in Fig.~\ref{QO10}b. Akin to the situation in Hg1223 $p$ = 0.112 shown in the main, the amplitude of the low frequency oscillation seen at $T$ = 4.2~K decreases as the temperature decreases, contrariwise to the behaviour expected by the Lifshitz-Kosevich theory\cite{Shoenberg}. As the low frequency gradually disappears, a higher frequency grows at low temperature but with a relatively smaller amplitude. This is clearly seen in the Fourier analysis of the oscillatory part of the data in Fig.~\ref{QO10}c. The amplitude of the peak at low frequency $F \approx $  650~T decreases as the temperature decreases and shift to slightly higher frequency. A peak at high frequency, $F$ = 2100~T,  emerges at low temperature.\\
Fig.~\ref{QOall} shows a comparison of the quantum oscillation spectrum in Hg1223 at different doping levels from $p$~=~0.08 to $p$~=~0.112. While the data looks qualitatively similar at $T$~=~4.2~K (see Fig.~\ref{QOall}a), the high frequency observed at low temperature occurs only for $p$ = 0.102 and $p$ = 0.112 (see Fig.~\ref{QOall}b).
%

\end{document}